\documentclass[a4paper,11pt]{article}
\usepackage{pos}
\usepackage{graphicx}

\title{Deciphering the Remnants of Core-Collapse Supernovae: Reconstructing Progenitor Star Properties and Explosion Mechanisms}
\ShortTitle{Deciphering the Remnants of Core-Collapse Supernovae}

\author*[a]{Salvatore Orlando}

\affiliation[a]{INAF - Osservatorio Astronomico di Palermo, Piazza
del Parlamento 1, 90134 Palermo, Italy}

\emailAdd{salvatore.orlando@inaf.it}

\abstract{Recent observations of the supernova remnant (SNR)
Cassiopeia A (Cas~A) with the James Webb Space Telescope (JWST)
have revealed unprecedented details of its structure, including an
intricate network of ejecta filaments and an enigmatic structure
known as the "Green Monster" (GM), characterized by a pockmarked
appearance with nearly circular holes and rings. These discoveries
offer unique insights into the mechanisms governing supernova (SN)
explosions and the subsequent evolution of ejecta and circumstellar
medium (CSM) interactions.

In this contribution, I present recent theoretical work based on
high-resolution three-dimensional (3D) hydrodynamic (HD) and magnetohydrodynamic (MHD)
simulations of a neutrino-driven SN explosion resembling Cas
A. The models follow the system from core collapse to an age of
$\sim 1000$ years, with the aim of investigating the origin and evolution
of these newly observed structures. The simulations incorporate key
physical processes, including neutrino-driven explosion dynamics,
hydrodynamic instabilities, Ni-bubble effects, and radiative losses,
while accounting for deviations from ionization equilibrium and
electron–proton temperature equilibration.

The new studies show that a web-like network of ejecta filaments, closely
resembling those observed by JWST, naturally forms during the early
stages of the explosion as a consequence of hydrodynamic instabilities
and the dynamic interplay of hot neutrino-driven bubbles. These
filaments preserve a "memory" of the initial explosion conditions
before being gradually disrupted by the reverse shock over centuries.
In parallel, the peculiar morphology of the GM can be reproduced
by the interaction of dense clumps and fingers of ejecta with an
asymmetric, forward-shocked CSM shell. Radiative cooling and
non-equilibrium ionization further enhance fragmentation, producing
dense knots and thin filaments that penetrate the shell and create
the observed network of holes and rings.

Our model highlights how the complex structures observed today in
Cas~A reflect both the imprint of the early SN explosion and the
later interactions of ejecta with the surrounding medium, offering
a unified view of the remnant's rich morphology.
}

\FullConference{Multifrequency Behaviour of High Energy Cosmic Sources - XV (MULTIF2025)\\
 9-14 June, 2025\\
Mondello, Palermo, Italy\\}

\begin{document}
\maketitle

\section{Introduction}

Supernova remnants (SNRs), the outcomes of supernova (SN) explosions,
can be essential laboratories for probing both the physics of
core-collapse SNe and the late evolutionary stages of massive stars.
When a massive star collapses, the explosion mechanism generates a
rich interplay of hydrodynamic instabilities, asymmetric plumes,
and turbulent mixing between nucleosynthetic layers
\cite{2003ApJ...584..971B, 2012ARNPS..62..407J, 2013A&A...552A.126W,
2015A&A...577A..48W, 2017hsn..book.1095J, 2018SSRv..214...33B,
2021Natur.589...29B}. These processes, occurring within the first
few seconds after core collapse, are usually inaccessible to direct
observation. Yet, as the stellar debris expands and interacts with
the circumstellar and interstellar environment, a SNR preserves
these signatures in its structure, composition, and dynamics
\cite{2016ApJ...822...22O, 2020A&A...636A..22O, 2021A&A...645A..66O,
2025A&A...699A.305O}. The geometry of the ejecta, the distribution
of heavy elements, and the morphology of shocked and unshocked
material all carry the imprint of the explosion engine. Studying
young SNRs therefore allows us to reconstruct aspects of the explosion
mechanism that cannot be directly inferred from SN light curves or
spectra.

At the same time, the outer morphology of a SNR, particularly the
shape and behavior of the forward and reverse shocks, also retains
information about the environment through which the remnant expands
\cite{2014ApJ...789....7L, 2021ApJ...922..140R, 2020ApJ...891..116W,
2024ApJ...966..147R, 2024MNRAS.532.3625M}. In the case of young
remnants, this environment is the circumstellar medium (CSM), which
reflects the mass-loss history of the progenitor star
\cite{2015ApJ...810..168O, 2022A&A...666A...2O, 2024ApJ...977..118O}.
Massive stars undergo complex evolutionary phases involving winds,
eruptions, and episodes of enhanced mass loss. These processes
create circumstellar shells, bubbles, and density gradients that
strongly influence the remnant's evolution once the SN blast wave
reaches them. As a result, young SNRs encapsulate both the physics
of the SN explosion and the final centuries or millennia of the
progenitor's life, bridging stellar evolution and SN physics.

Among all Galactic remnants, Cassiopeia A (Cas~A) stands as the
most informative example of this dual role. With an age of only
$\approx 350$ years \cite{2001AJ....122..297T, 2006ApJ...645..283F},
Cas~A is one of the few SNRs young enough that many fine-scale
structures originating from the explosion are still visible. Its
proximity ($\approx 3.4$ kpc, \cite{1995ApJ...440..706R}) allows
high-resolution observations across the electromagnetic spectrum,
revealing a wealth of morphological features \cite{2003ApJ...597..347L,
2010ApJ...725.2038D, 2012ApJ...746..130H, 2013ApJ...772..134M,
2014Natur.506..339G, 2015Sci...347..526M, 2020ApJ...895...82V,
2020ApJ...889..144H, 2025ApJ...990L...5S}: Fe-rich plumes that
extend beyond lighter elements, an inversion of nucleosynthetic
layers, rapidly moving knots and jets, and strong asymmetries in
both shocked and unshocked material. These characteristics have
made Cas~A the benchmark case for testing multidimensional SN models
and for constraining the nature of mixing instabilities, neutrino-driven
convection, and asymmetries in the explosion engine.

The advent of JWST has profoundly enhanced this picture. Infrared
imaging of Cas~A has unveiled a complex network of unshocked, O-rich
filaments in its interior \cite{2024ApJ...965L..27M, 2024ApJ...969L...9R},
exhibiting sub-parsec structures with extraordinary sharpness.
Additionally, JWST observations of the near (blueshifted) side of
the remnant have revealed the striking "Green Monster"
\cite{2024ApJ...976L...4D} (GM): a region filled with circular
holes, partial rings, and arc-like cavities in shocked circumstellar
material. These discoveries pose new challenges: the internal
filaments require an explanation rooted in the earliest instabilities
of the explosion, while the GM demands a mechanism involving the
interaction between ejecta clumps and the progenitor's circumstellar
environment.

Two recent numerical works, one dedicated to the origin of the
filamentary interior network \cite{2025A&A...696A.108O}, and the
other to the formation of the GM's holes and rings
\cite{2025A&A...696A.188O}, address these challenges with fully
three-dimensional (3D) simulations following Cas~A's evolution from
the core-collapse to the present day and, extending the evolution
to the age of 1000 years \cite{2021A&A...645A..66O}. Together, these
studies aim to link the imprints of the explosion engine, the
behavior of the ejecta, and the structure of the CSM into a coherent
narrative. The present review summarizes the main results.

\section{Modeling Strategy}

The two studies build upon a unified numerical framework designed
to follow the evolution of Cas~A continuously from the first instants
after core collapse to the present remnant stage (see Orlando et
al. \cite{2021A&A...645A..66O} for the details of the simulations).
The modeling begins with a 3D neutrino-driven explosion simulation,
specifically the W15-2-cw-IIb model \cite{2017ApJ...842...13W},
which provides the structure of the ejecta only hours after the
launch of the outgoing shock. This model is based on a $15\,M_\odot$
zero-age main sequence progenitor that was artificially stripped
of most of its hydrogen envelope to reproduce a Type~IIb configuration
appropriate for Cas~A, resulting in a compact pre-supernova star
with a final radius of $\sim 21.5\,R_\odot$. The explosion was
calibrated to reach a final kinetic energy of $\sim 1.5\times10^{51}$~erg
with a total mass of ejecta of $3.3\,M_\odot$, and to synthesize
approximately $\approx 0.1\,M_\odot$ of radioactive $^{56}$Ni. Only
about $0.3\,M_\odot$ of hydrogen-rich material was
retained in the progenitor structure, implying that roughly
$8-10\,M_\odot$ of stellar mass was lost during the late evolutionary
stages and injected into the surrounding CSM. At
this early time, the explosion already contains the key physical
ingredients that govern the later morphology of the remnant:
large-scale neutrino-heated bubbles, strong Rayleigh-Taylor (RT)
instabilities, and high-velocity Ni-rich plumes punching through
the star's composition layers. These features break spherical
symmetry from the outset and establish a network of density contrasts,
plumes, and cavities that later evolve into the filaments and clumps
observed in the remnant.

This initial configuration is mapped to a large computational grid
and evolved for centuries using high-resolution 3D hydrodynamics
or magnetohydrodynamics (MHD) when appropriate. The simulations
track the free expansion of the ejecta, the formation of the reverse
shock, and the progressive interaction with the surrounding CSM.
They incorporate a comprehensive set of physical processes essential
to capturing the behavior of a young SNR: optically thin radiative
cooling, which allows dense ejecta knots to condense and survive;
non-equilibrium ionization, which governs the thermal state of both
shocked and unshocked gas; and heating from radioactive decay, in
particular the expansion of Ni-rich bubbles that compress O-rich
layers. Chemical elements are followed with passive tracers, enabling
the reconstruction of spatial distributions for individual species
such as O, Si, Fe, and Ti. Thanks to a remapping technique (e.g.,
Orlando et al. \cite{2019A&A...622A..73O, 2020A&A...636A..22O}) and
high spatial resolution ($2048^3$ cells), the simulations resolve
structures down to $\sim 0.01$~pc at the age of Cas~A, comparable
to the spatial resolution of JWST.

To investigate the GM, the evolving ejecta interact with
a thin, dense circumstellar shell positioned at roughly 1.5 pc from
the explosion center. This shell, previously proposed to explain
Cas~A's reverse-shock asymmetry \cite{2022A&A...666A...2O}, is
modeled as a narrow, high-density layer, likely produced by a major
mass-loss episode of the progenitor star shortly ($10^4-10^5$ years)
before core collapse. Its structure is asymmetric, with enhanced
density on the blueshifted side of the remnant (the region where
the GM is observed). This shell plays a crucial role in
shaping the late-time morphology, acting as a barrier that is struck
and perforated by the fastest ejecta clumps.

Finally, the simulations are rotated into the orientation that best
matches the observed distribution of Fe- and Ti-rich ejecta in Cas
A. Synthetic projections of density, temperature, and element-specific
emission allow a direct comparison with JWST, Chandra, and NuSTAR
observations. Through this combination of realistic explosion
physics, detailed microphysics, and careful observational alignment,
the models provide a physically grounded, high-fidelity reconstruction
of Cas~A's evolution from core collapse to its present, highly
structured remnant state.

\section{Results}

To understand the new structures revealed in Cas~A by JWST, the
simulations examine different but complementary aspects of the
remnant's evolution. Sect.~\ref{sec:filaments} focuses on the
internal, unshocked O-rich filamentary network, tracing its origin
back to the earliest seconds of the explosion, while
Sect.~\ref{sec:Gmonster} investigates the GM, showing how its system
of holes and rings arises from the interaction between ejecta clumps
and a dense circumstellar shell. Together, these works
\cite{2025A&A...696A.108O, 2025A&A...696A.188O} provide a coherent
physical interpretation linking the explosion mechanism, the evolution
of the ejecta, and the imprint of the progenitor's mass-loss history.

\subsection{Origin of the Filamentary O-Rich Network}
\label{sec:filaments}

The simulations demonstrate that the intricate web of unshocked,
O-rich filaments revealed by JWST in the interior of Cas~A is not
a product of late-time hydrodynamic evolution in the SNR, but instead
a direct fossil imprint of the explosion mechanism itself, established
within the first second after core collapse \cite{2025A&A...696A.108O}.
In the model, the neutrino-driven engine produces large, rising
plumes of neutrino-heated material that immediately break spherical
symmetry. As the shock moves outward through the stratified stellar
layers, these plumes drive vigorous mixing across nucleosynthetic
boundaries.

\begin{figure}
  \begin{center}
    \includegraphics[width=5.5cm]{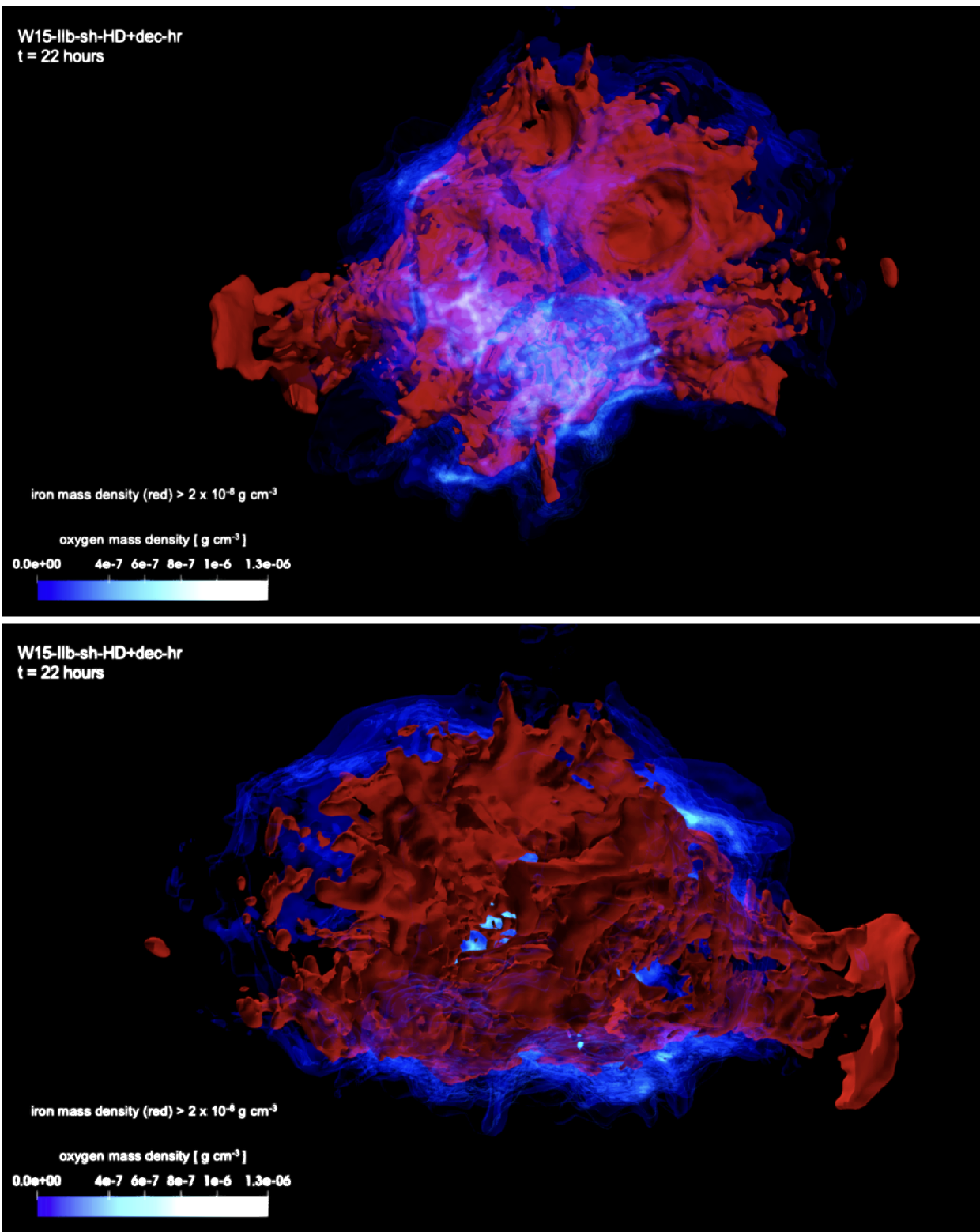}
      \caption{Distribution of unshocked Ni-rich ejecta (red
      isosurface; $\rho_{\mathrm{Ni}} > 2 \times
      10^{-8}\,\mathrm{g\,cm^{-3}}$) a few hours after shock breakout
      ($\approx 22$ hr after core collapse) for model
      \texttt{W15-IIb-sh-HD+dec-hr}. Unshocked O-rich ejecta are
      shown via volume rendering (blue palette). The upper left
      panel shows the front view as seen from Earth. The remaining
      panels show perspectives from arbitrary orientations. An
      interactive 3D visualization of the O and Ni distributions
      is available at \url{https://skfb.ly/psXKs}.}
  \label{fig1}
  \end{center}
\end{figure}

In the first minutes to hours after shock launch, the formation and
growth of hot, buoyant, Ni-rich bubbles further amplify the initial
asymmetries. Their expansion pushes and compresses the surrounding
layers enriched in oxygen, carbon, neon, and magnesium, carving
them into elongated sheets. At the same time, RT instabilities
rapidly stretch, corrugate, and fragment these dense structures,
producing a complex filamentary pattern at the time when the shock
reaches the stellar surface (see Fig.~\ref{fig1}).

In the following months, radioactive decay of Ni to Co and later
to Fe releases additional energy that inflates the originally Ni-rich
ejecta, an effect known as the Ni-bubble phenomenon. This continued
expansion reinforces the compression of the surrounding intermediate-mass
layers, sharpening the internal structure and further refining the
filamentary network. As a result, by the time the remnant is only
a few years old, the simulations already display a highly organized,
interconnected web of dense, O-rich filaments occupying the remnant's
unshocked interior.

\begin{figure}
  \begin{center}
    \includegraphics[width=5cm]{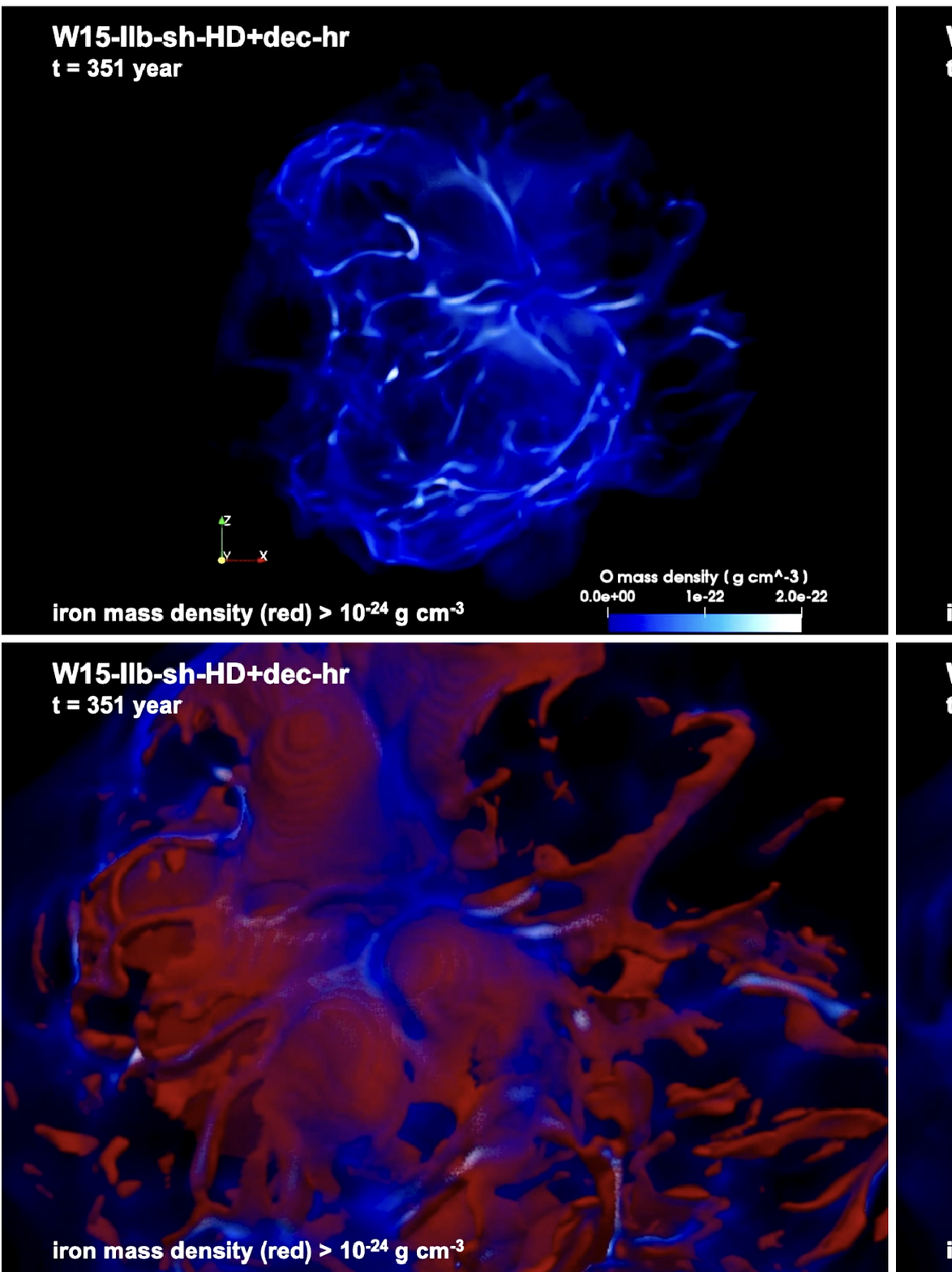}
      \caption{Upper panels: Distribution of unshocked O-rich ejecta
      visualized via volume rendering (blue palette) at the age of
      Cas~A for model \texttt{W15-IIb-sh-HD+dec-hr}. The color scale
      is shown in the bottom-right corner of each panel. Rendering
      opacity is proportional to the plasma density, highlighting
      the denser structures. The upper-left panel shows the front
      view as seen from Earth, while the upper-right panel shows a
      side view from a vantage point to the west (positive $x$-axis).
      Lower panels: Zoom into the central region revealing filamentary
      O-rich ejecta. In the lower-left panel, the O-rich distribution
      (blue) is shown together with the unshocked Fe-rich ejecta,
      displayed as a red isosurface corresponding to Fe densities
      above $10^{-24}\,\mathrm{g\,cm^{-3}}$. An interactive 3D
      visualization of the O and Fe spatial distributions at the
      age of Cas~A is available at \url{https://skfb.ly/psXKr}.}
  \label{fig2}
  \end{center}
\end{figure}

At the current age of $\sim 350$ years, the filamentary network in
the models exhibits an analogous geometry, hierarchy, and characteristic
thickness ($\sim 0.01$ pc) as JWST images (see Fig.~\ref{fig2}): a
tangled internal "spider-web" of O-rich structures occupying the
interior of the remnant, bounded by the cavities created by fast-rising
Fe-rich plumes. Heavier material such as Fe and Ti remains largely
concentrated within these plumes and cavities and does not participate
in the extended filamentary web. The spatial anticorrelation between
shocked Fe-rich structures and O-rich filaments thus emerges naturally
from the explosion physics, without requiring fine-tuned assumptions
in the remnant phase (see Orlando et al. \cite{2025A&A...696A.108O}
for the details).

The simulations also roughly reproduce the observed distribution
of unshocked mass: at Cas~A's present epoch, roughly one solar mass
of ejecta remains unprocessed by the reverse shock, consistent with
observational estimates. The unshocked component is dominated by
oxygen, with smaller contributions from helium, carbon, neon,
magnesium, and silicon, and only trace amounts of Fe-group nuclei.
These masses and abundances fall within observational constraints
from infrared and X-ray studies.

Once the reverse shock encounters the filamentary ejecta, beginning
a few decades after explosion, the network starts to degrade. RT
and Kelvin–Helmholtz instabilities generated by the shock interaction
gradually erode the fine-scale structure, fragmenting the filaments
and dispersing their material into the shocked interior. The modeling
predicts that although the network is still prominently visible
today, it is transient: as Cas~A approaches $\sim 700$ years of
age, the filaments lose coherence and the JWST-scale morphology
will no longer be recognizable. We note that this also implies that
similar filament networks in older Galactic remnants would have
been erased, making Cas~A a uniquely timed system for observing the
explosion's hydrodynamic fingerprint.

Finally, the study emphasizes that the formation of the filamentary
network depends predominantly on the physics of the SN explosion,
not on the properties of the CSM or later remnant
evolution. Even in simulations lacking strong circumstellar shaping,
the network forms, evolves, and reaches a structure in remarkable
agreement with observations. Thus, JWST's newly revealed filaments
provide the clearest evidence to date that young SNRs can function
as direct archaeological records of the earliest instabilities of
the explosion engine, allowing modern simulations and observations
to probe the inner seconds of a core-collapse event centuries after
it occurred.

\subsection{Origin of the Green Monster}
\label{sec:Gmonster}

\begin{figure}
  \begin{center}
    \includegraphics[width=15cm]{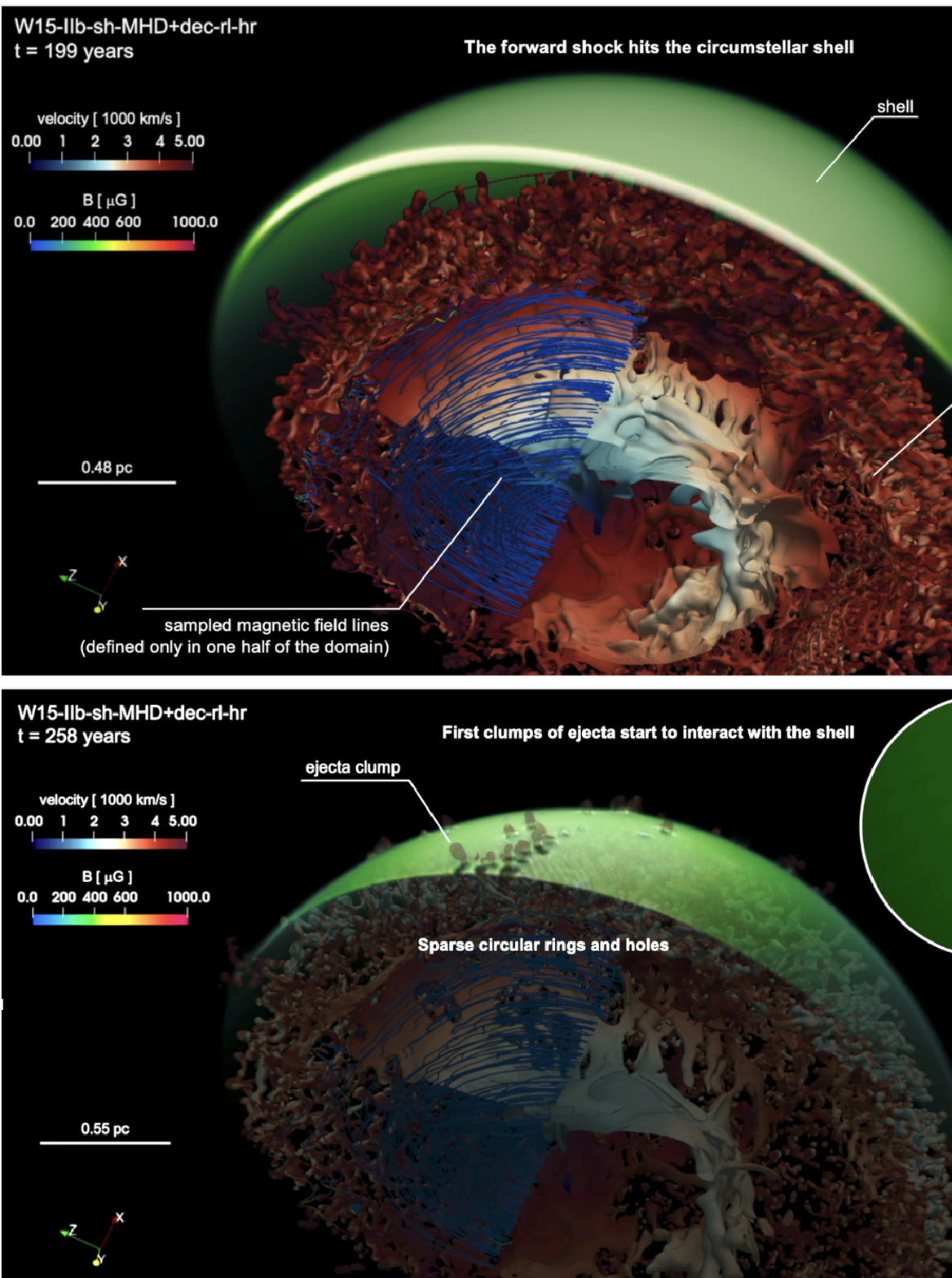}
       \caption{3D visualization of the ejecta-shell interaction
       in the Cas~A SNR from model \texttt{W15-IIb-sh-MHD+dec-rl-hr}. The
       irregular isosurface represents ejecta with mass density
       exceeding $10^{-23}\,\mathrm{g\,cm^{-3}}$, with colors
       indicating the radial velocity in units of
       $1000\,\mathrm{km\,s^{-1}}$ (color scale shown in the left
       panels). The green volume rendering depicts the mass density
       of the shocked shell material. The sequence spans the evolution
       from the moment the forward shock first encounters the shell
       at $t \approx 200$~years (upper-left panel) to the later remnant
       stage at $t \approx 1000$~years (lower-right panel). The panels
       on the right show the remnant–shell interaction at three
       representative epochs, with times indicated in the upper-left
       corner of each panel. The upper-right inset provides a
       detailed view of the shell structure, showing the progressive
       development of holes and ring-like features produced by the
       interaction with the ejecta. For clarity, magnetic field
       lines (color scale shown in the left panels) are shown only
       within a selected sub-volume, highlighting the complexity
       of the magnetic configuration while preserving a clear view
       of the ejecta structure in other regions of the remnant.
       Interactive 3D visualizations of the remnant-shell interaction
       are available at \url{https://skfb.ly/psYpK} and
       \url{https://skfb.ly/pt7wt}.}
  \label{fig3}
  \end{center}
\end{figure}

The simulations also show that the GM (the large, pockmarked,
ring-filled region identified by JWST on the blueshifted, near side
of Cas~A) arises naturally from the interaction of fast, clumpy SN
ejecta with a thin, dense circumstellar shell produced by the
progenitor star before explosion \cite{2025A&A...696A.188O}. In the
simulations, this shell lies at a radius of about 1.5 pc, with a
thickness of only $\sim 0.02$ pc and enhanced density in the northwest
direction (see Orlando et al. \cite{2022A&A...666A...2O} for more
details), matching roughly the observed location of the GM. When
the ejecta, shaped by early mixing instabilities in the SN,
expand into this structure, dense radiatively cooled RT fingers of
ejecta collide with the shell at several thousand kilometers per
second, leading to highly characteristic morphological signatures
that directly correspond to the JWST data. Figure~\ref{fig3} shows
the different phases during the interaction of the remnant with the
shell and the formation of the holes and rings.

\begin{figure}
  \begin{center}
    \includegraphics[width=15cm]{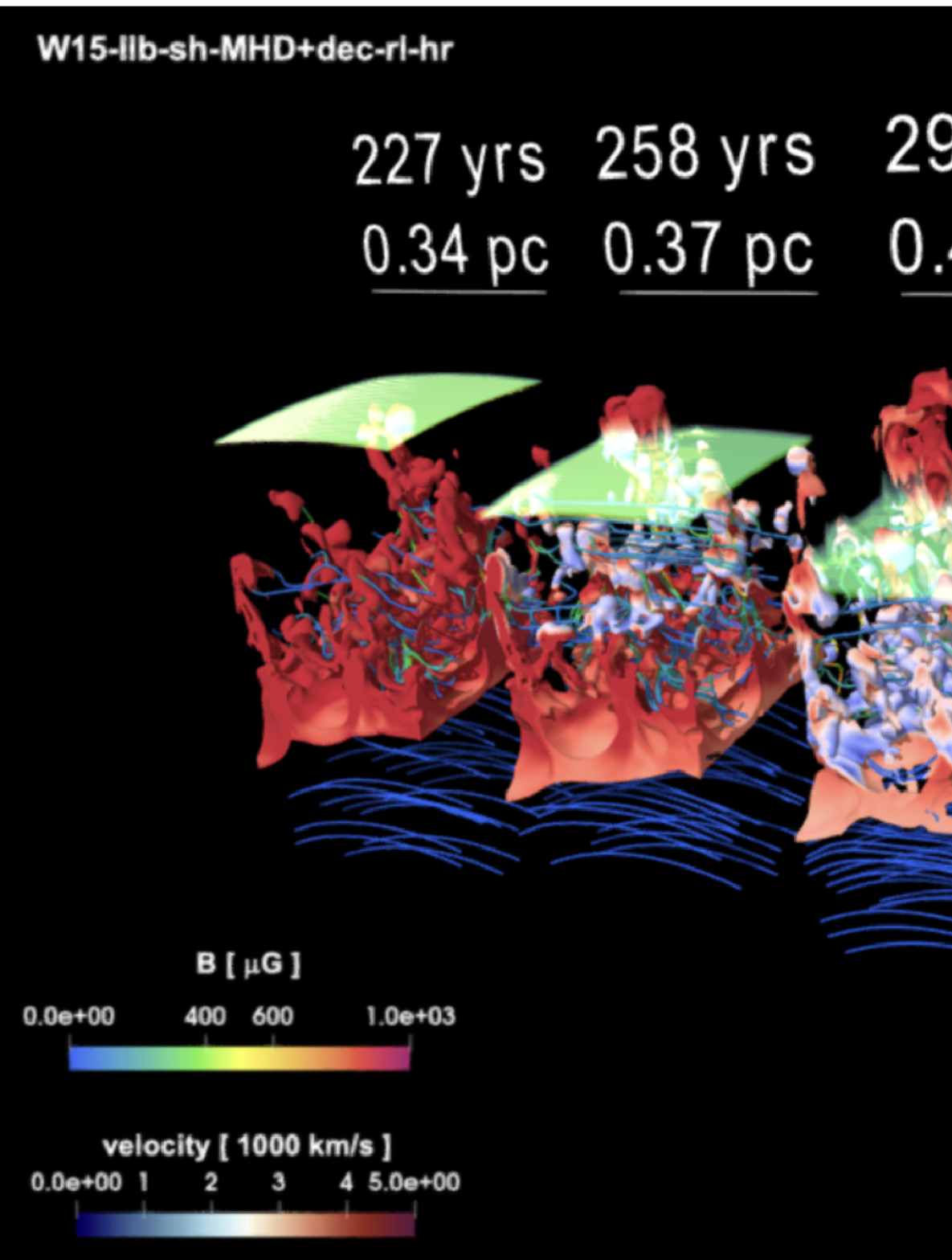}
      \caption{Formation of holes and ring-like structures during
      the interaction between the ejecta and the shocked shell in
      model \texttt{W15-IIb-sh-MHD+dec-rl-hr}. The irregular
      isosurface traces the ejecta, while the green volume rendering
      shows the mass density of the shocked shell material, as
      illustrated in Fig.~\ref{fig3}. The sequence follows the evolution
      from the moment the ejecta fingers first encounter the shell
      at $t \approx 227$~years (left) to a more advanced remnant
      stage at $t \approx 367$~years (right), corresponding to the
      age of Cas~A. An interactive 3D visualization of the ejecta
      fingers protruding through the shell is available at
      \url{https://skfb.ly/ptFHY}.}
  \label{fig4}
  \end{center}
\end{figure}

In the simulations, the dense RT fingers produced by the explosion
behave effectively as ballistic streams as they encounter the
circumstellar shell. Their high velocity allows them to punch through
the shell, opening narrow perforations in its structure. The material
displaced by the impact is swept laterally and accumulates around
the penetration site, forming near-circular rims that outline the
resulting holes. Figure~\ref{fig4} illustrates the details of this
interaction for a small portion of the remnant shell interaction.
When several RT fingers strike the shell in close
proximity, their effects overlap, producing compound, irregular,
or partially merged rings that reproduce the complex "Swiss-cheese"
appearance characteristic of the GM region observed by JWST.

Radiative cooling plays a decisive role, allowing the fingers to
reach extremely high density contrasts relative to the shell and
remain thin and coherent long enough to create clean perforations
with a size comparable to that observed rather than dispersing upon
impact. In models that include magnetic fields, the knots are
confined even more efficiently, strengthening their ability to
survive and burrow through the shell. The simulations also show
that the appearance of the GM is dominated not by optical depth or
viewing-angle effects but by real physical cavities created in the
shocked circumstellar material.

The sizes of the simulated holes (typically around 5 arcseconds,
with a distribution extending from $\sim 1$ to $>10$ arcseconds)
match well with JWST measurements. A minority of synthetic holes
are larger than observed and could be due to the circumstellar shell
being located slightly farther from the explosion center in the
actual system or possessing a more complex structure than represented
in the models. The chemical composition inside the holes is heavily
enriched in O-, C-, and Ne-rich material, reflecting the intermediate-mass
layers of the progenitor star and providing clear predictions for
future spectroscopic confirmation.

\section{Conclusions}

The recent simulations describing Cas~A demonstrate the power of
following the evolution of a core-collapse explosion in a fully
self-consistent manner: from the progenitor star through the SN
event and into the SNR phase. The studies discussed
here show that many of the most striking morphological features
revealed by JWST in Cas~A cannot be understood if the SNR is modeled
without connecting the different phases of evolution, from the
progenitor star to the SNR. Instead, they arise naturally from
physical processes occurring much earlier: minutes, hours, or even
seconds after core collapse for the filaments shaping the remnant’s
interior, and centuries to millennia before the SN event in
the case of the GM. The filamentary O-rich network is a fossilized
record of the neutrino-driven instabilities and mixing processes
that shaped the explosion engine, while the GM owes its
origin to the interaction of the resulting ejecta structures with
a circumstellar shell created by the late stages of the progenitor's
evolution. These features therefore encapsulate, within a single
remnant, the history of both the explosion mechanism and the mass-loss
behavior of the star that produced it.

These studies highlight the need for models that seamlessly bridge
stellar evolution, SN dynamics, and long-term remnant
development. For decades, theoretical studies have tended to treat
these phases separately, largely because of the complexity of
simulating such a wide range of time and spatial scales. However,
the unprecedented sensitivity and angular resolution of instruments
such as JWST and future facilities now reveal structures that
directly connect these phases in observable detail. Under these
circumstances, it becomes essential to model the entire evolutionary
chain in a unified framework capable of tracking how instabilities
seeded in the explosion produce observable signatures centuries
later, and how the circumstellar environment imprinted by the
progenitor star shapes the propagation and appearance of the ejecta.

Self-consistent models of this kind do not merely reproduce images;
they allow young SNRs to function as archaeological records of the
explosion engine, preserving the fingerprints of neutrino heating,
hydrodynamic instabilities, and radioactive energy deposition. At
the same time, they probe the final centuries of stellar evolution,
revealing the density, geometry, and asymmetries of late stellar
winds and eruptive mass loss. Cas~A is currently the most powerful
demonstration of this approach, but as new observatories continue
to discover and resolve young remnants across our Galaxy and the
Local Group, similar studies will become increasingly important.
Another remarkable example is SN~1987A, whose ongoing interaction
with its inhomogeneous CSM provides an unparalleled laboratory
for testing explosion models, progenitor mass-loss histories, and
the transition from SN to SNR. Recent observations are now revealing
previously inaccessible details of the ejecta structure in SN~1987A
\cite{2013ApJ...768...89L, 2016ApJ...833..147L, 2023ApJ...958...95J,
2023ApJ...949L..27L, 2024Sci...383..898F, 2024MNRAS.532.3625M,
2025ApJ...981...26S}, providing the crucial spatial, chemical, and
dynamical information needed to identify the underlying explosion
mechanism and to constrain the nature of the progenitor star
\cite{2020A&A...636A..22O, 2021ApJ...908L..45G, 2022ApJ...931..132G,
2024ApJ...961L...9S, 2025A&A...699A.305O, 2025PASJ...77S.193X}.

In this context, the synergy between high-fidelity simulations and
high-quality multiwavelength observations marks a new era in the
study of massive stars and core-collapse SNe. By linking the
physics of the progenitor, the explosion, and the remnant in a
single, continuous evolutionary narrative, we can now extract
physical information that was previously inaccessible, constraining
the operation of the SN engine, the role of neutrino-driven
turbulence, the development of mixing and RT instabilities, and the
nature of massive star mass loss immediately before collapse. This
approach promises to transform young remnants like Cas~A (and SN
1987A) into precise diagnostic tools, enabling us to move beyond
phenomenology and toward a deeper physical understanding of how
massive stars end their lives.

\acknowledgments
We thank an anonymous referee for their suggestions that improved
our paper. Many colleagues have contributed to the findings reported
here; in particular, I am grateful to H.-T. Janka and A. Wongwathanarat
for providing the supernova explosion model, and to D. Milisavljevic
for leading the analysis of the JWST observations. Additional key
contributions were provided by (in alphabetical order) F. Bocchino,
I. De Looze, D. Dickinson, E. Greco, M. Miceli, S. Nagataki, M.
Ono, D. Patnaude, J. Rho, V. Sapienza, and T. Temim, whose efforts
were fundamental to the success of the project. I acknowledge partial
financial contribution from the PRIN 2022 (20224MNC5A) - ``Life,
death and after-death of massive stars'' funded by European Union
– Next Generation EU and from the INAF Theory Grant ``Supernova
remnants as probes for the structure and mass-loss history of the
progenitor systems''. The navigable 3D graphics have been developed
in the framework of the project 3DMAP-VR (3-Dimensional Modeling
of Astrophysical Phenomena in Virtual Reality; \cite{2019RNAAS...3..176O,
2023MmSAI..94a..13O}) at INAF-Osservatorio Astronomico di Palermo.

\bibliographystyle{aa}
\bibliography{references}

\begin{thebibliography}{50}
\expandafter\ifx\csname natexlab\endcsname\relax\def\natexlab#1{#1}\fi

\bibitem[{{Blondin} {et~al.}(2003){Blondin}, {Mezzacappa}, \&
  {DeMarino}}]{2003ApJ...584..971B}
{Blondin}, J.~M., {Mezzacappa}, A., \& {DeMarino}, C. 2003, \apj, 584, 971

\bibitem[{{Burrows} \& {Vartanyan}(2021)}]{2021Natur.589...29B}
{Burrows}, A. \& {Vartanyan}, D. 2021, \nat, 589, 29

\bibitem[{{Burrows} {et~al.}(2018){Burrows}, {Vartanyan}, {Dolence}, {Skinner},
  \& {Radice}}]{2018SSRv..214...33B}
{Burrows}, A., {Vartanyan}, D., {Dolence}, J.~C., {Skinner}, M.~A., \&
  {Radice}, D. 2018, \ssr, 214, 33

\bibitem[{{De Looze} {et~al.}(2024){De Looze}, {Milisavljevic}, {Temim},
  {Dickinson}, {Fesen}, {Arendt}, {Chastenet}, {Orlando}, {Vink}, {Barlow},
  {Kirchschlager}, {Priestley}, {Raymond}, {Rho}, {Sartorio}, {Scheffler},
  {Schmidt}, {Blair}, {Fox}, {Fryer}, {Janka}, {Koo}, {Laming}, {Matsuura},
  {Patnaude}, {Rela{\~n}o}, {Rest}, {Schmidt}, {Smith}, \&
  {Sravan}}]{2024ApJ...976L...4D}
{De Looze}, I., {Milisavljevic}, D., {Temim}, T., {et~al.} 2024, \apjl, 976, L4

\bibitem[{{DeLaney} {et~al.}(2010){DeLaney}, {Rudnick}, {Stage}, {Smith},
  {Isensee}, {Rho}, {Allen}, {Gomez}, {Kozasa}, {Reach}, {Davis}, \&
  {Houck}}]{2010ApJ...725.2038D}
{DeLaney}, T., {Rudnick}, L., {Stage}, M.~D., {et~al.} 2010, \apj, 725, 2038

\bibitem[{{Fesen} {et~al.}(2006){Fesen}, {Hammell}, {Morse}, {Chevalier},
  {Borkowski}, {Dopita}, {Gerardy}, {Lawrence}, {Raymond}, \& {van den
  Bergh}}]{2006ApJ...645..283F}
{Fesen}, R.~A., {Hammell}, M.~C., {Morse}, J., {et~al.} 2006, \apj, 645, 283

\bibitem[{{Fransson} {et~al.}(2024){Fransson}, {Barlow}, {Kavanagh}, {Larsson},
  {Jones}, {Sargent}, {Meixner}, {Bouchet}, {Temim}, {Wright}, {Blommaert},
  {Habel}, {Hirschauer}, {Hjorth}, {Lenki{\'c}}, {Tikkanen}, {Wesson},
  {Coulais}, {Fox}, {Gastaud}, {Glasse}, {Jaspers}, {Krause}, {Lau}, {Nayak},
  {Rest}, {Colina}, {van Dishoeck}, {G{\"u}del}, {Henning}, {Lagage},
  {{\"O}stlin}, {Ray}, \& {Vandenbussche}}]{2024Sci...383..898F}
{Fransson}, C., {Barlow}, M.~J., {Kavanagh}, P.~J., {et~al.} 2024, Science,
  383, 898

\bibitem[{{Greco} {et~al.}(2021){Greco}, {Miceli}, {Orlando}, {Olmi},
  {Bocchino}, {Nagataki}, {Ono}, {Dohi}, \& {Peres}}]{2021ApJ...908L..45G}
{Greco}, E., {Miceli}, M., {Orlando}, S., {et~al.} 2021, \apjl, 908, L45

\bibitem[{{Greco} {et~al.}(2022){Greco}, {Miceli}, {Orlando}, {Olmi},
  {Bocchino}, {Nagataki}, {Sun}, {Vink}, {Sapienza}, {Ono}, {Dohi}, \&
  {Peres}}]{2022ApJ...931..132G}
{Greco}, E., {Miceli}, M., {Orlando}, S., {et~al.} 2022, \apj, 931, 132

\bibitem[{{Grefenstette} {et~al.}(2014){Grefenstette}, {Harrison}, {Boggs},
  {Reynolds}, {Fryer}, {Madsen}, {Wik}, {Zoglauer}, {Ellinger}, {Alexand er},
  {An}, {Barret}, {Christensen}, {Craig}, {Forster}, {Giommi}, {Hailey},
  {Hornstrup}, {Kaspi}, {Kitaguchi}, {Koglin}, {Mao}, {Miyasaka}, {Mori},
  {Perri}, {Pivovaroff}, {Puccetti}, {Rana}, {Stern}, {Westergaard}, \&
  {Zhang}}]{2014Natur.506..339G}
{Grefenstette}, B.~W., {Harrison}, F.~A., {Boggs}, S.~E., {et~al.} 2014, \nat,
  506, 339

\bibitem[{{Holland-Ashford} {et~al.}(2020){Holland-Ashford}, {Lopez}, \&
  {Auchettl}}]{2020ApJ...889..144H}
{Holland-Ashford}, T., {Lopez}, L.~A., \& {Auchettl}, K. 2020, \apj, 889, 144

\bibitem[{{Hwang} \& {Laming}(2012)}]{2012ApJ...746..130H}
{Hwang}, U. \& {Laming}, J.~M. 2012, \apj, 746, 130

\bibitem[{{Janka}(2012)}]{2012ARNPS..62..407J}
{Janka}, H.-T. 2012, Annual Review of Nuclear and Particle Science, 62, 407

\bibitem[{{Janka}(2017)}]{2017hsn..book.1095J}
{Janka}, H.-T. 2017, {``Neutrino-Driven Explosions'' chapter in {\it Handbook
  of Supernovae}} (edited by Athem W. Alsabti and Paul Murdin, ISBN
  978-3-319-21845-8. Springer International Publishing, Switzerland), p.~1095

\bibitem[{{Jones} {et~al.}(2023){Jones}, {Kavanagh}, {Barlow}, {Temim},
  {Fransson}, {Larsson}, {Blommaert}, {Meixner}, {Lau}, {Sargent}, {Bouchet},
  {Hjorth}, {Wright}, {Coulais}, {Fox}, {Gastaud}, {Glasse}, {Habel},
  {Hirschauer}, {Jaspers}, {Krause}, {Lenki{\'c}}, {Nayak}, {Rest}, {Tikkanen},
  {Wesson}, {Colina}, {van Dishoeck}, {G{\"u}del}, {Henning}, {Lagage},
  {{\"O}stlin}, {Ray}, \& {Vandenbussche}}]{2023ApJ...958...95J}
{Jones}, O.~C., {Kavanagh}, P.~J., {Barlow}, M.~J., {et~al.} 2023, \apj, 958,
  95

\bibitem[{{Laming} \& {Hwang}(2003)}]{2003ApJ...597..347L}
{Laming}, J.~M. \& {Hwang}, U. 2003, \apj, 597, 347

\bibitem[{{Larsson} {et~al.}(2013){Larsson}, {Fransson}, {Kjaer}, {Jerkstrand},
  {Kirshner}, {Leibundgut}, {Lundqvist}, {Mattila}, {McCray}, {Sollerman},
  {Spyromilio}, \& {Wheeler}}]{2013ApJ...768...89L}
{Larsson}, J., {Fransson}, C., {Kjaer}, K., {et~al.} 2013, \apj, 768, 89

\bibitem[{{Larsson} {et~al.}(2023){Larsson}, {Fransson}, {Sargent}, {Jones},
  {Barlow}, {Bouchet}, {Meixner}, {Blommaert}, {Coulais}, {Fox}, {Gastaud},
  {Glasse}, {Habel}, {Hirschauer}, {Hjorth}, {Jaspers}, {Kavanagh}, {Krause},
  {Lau}, {Lenki{\'c}}, {Nayak}, {Rest}, {Temim}, {Tikkanen}, {Wesson}, \&
  {Wright}}]{2023ApJ...949L..27L}
{Larsson}, J., {Fransson}, C., {Sargent}, B., {et~al.} 2023, \apjl, 949, L27

\bibitem[{{Larsson} {et~al.}(2016){Larsson}, {Fransson}, {Spyromilio},
  {Leibundgut}, {Challis}, {Chevalier}, {France}, {Jerkstrand}, {Kirshner},
  {Lundqvist}, {Matsuura}, {McCray}, {Smith}, {Sollerman}, {Garnavich}, {Heng},
  {Lawrence}, {Mattila}, {Migotto}, {Sonneborn}, {Taddia}, \&
  {Wheeler}}]{2016ApJ...833..147L}
{Larsson}, J., {Fransson}, C., {Spyromilio}, J., {et~al.} 2016, \apj, 833, 147

\bibitem[{{Lee} {et~al.}(2014){Lee}, {Park}, {Hughes}, \&
  {Slane}}]{2014ApJ...789....7L}
{Lee}, J.-J., {Park}, S., {Hughes}, J.~P., \& {Slane}, P.~O. 2014, \apj, 789, 7

\bibitem[{{Matsuura} {et~al.}(2024){Matsuura}, {Boyer}, {Arendt}, {Larsson},
  {Fransson}, {Rest}, {Ravi}, {Park}, {Cigan}, {Temim}, {Dwek}, {Barlow},
  {Bouchet}, {Clayton}, {Chevalier}, {Danziger}, {De Buizer}, {De Looze}, {De
  Marchi}, {Fox}, {Gall}, {Gehrz}, {Gomez}, {Indebetouw}, {Kangas},
  {Kirchschlager}, {Kirshner}, {Lundqvist}, {Marcaide}, {Mart{\'\i}-Vidal},
  {Meixner}, {Milisavljevic}, {Orlando}, {Otsuka}, {Priestley}, {Richards},
  {Schmidt}, {Staveley-Smith}, {Smith}, {Spyromilio}, {Vink}, {Wang}, {Watson},
  {Wesson}, {Wheeler}, {Woodward}, {Zanardo}, {Alp}, \&
  {Burrows}}]{2024MNRAS.532.3625M}
{Matsuura}, M., {Boyer}, M., {Arendt}, R.~G., {et~al.} 2024, \mnras, 532, 3625

\bibitem[{{Milisavljevic} \& {Fesen}(2013)}]{2013ApJ...772..134M}
{Milisavljevic}, D. \& {Fesen}, R.~A. 2013, \apj, 772, 134

\bibitem[{{Milisavljevic} \& {Fesen}(2015)}]{2015Sci...347..526M}
{Milisavljevic}, D. \& {Fesen}, R.~A. 2015, Science, 347, 526

\bibitem[{{Milisavljevic} {et~al.}(2024){Milisavljevic}, {Temim}, {De Looze},
  {Dickinson}, {Laming}, {Fesen}, {Raymond}, {Arendt}, {Vink}, {Posselt},
  {Pavlov}, {Fox}, {Pinarski}, {Subrayan}, {Schmidt}, {Blair}, {Rest},
  {Patnaude}, {Koo}, {Rho}, {Orlando}, {Janka}, {Andrews}, {Barlow}, {Burrows},
  {Chevalier}, {Clayton}, {Fransson}, {Fryer}, {Gomez}, {Kirchschlager}, {Lee},
  {Matsuura}, {Niculescu-Duvaz}, {Pierel}, {Plucinsky}, {Priestley}, {Ravi},
  {Sartorio}, {Schmidt}, {Shahbandeh}, {Slane}, {Smith}, {Sravan}, {Weil},
  {Wesson}, \& {Wheeler}}]{2024ApJ...965L..27M}
{Milisavljevic}, D., {Temim}, T., {De Looze}, I., {et~al.} 2024, \apjl, 965,
  L27

\bibitem[{{Orlando} {et~al.}(2024){Orlando}, {Greco}, {Hirai}, {Matsuoka},
  {Miceli}, {Nagataki}, {Ono}, {Chen}, {Milisavljevic}, {Patnaude}, {Bocchino},
  \& {Elias-Rosa}}]{2024ApJ...977..118O}
{Orlando}, S., {Greco}, E., {Hirai}, R., {et~al.} 2024, \apj, 977, 118

\bibitem[{{Orlando} {et~al.}(2025{\natexlab{a}}){Orlando}, {Janka},
  {Wongwathanarat}, {Bocchino}, {De Looze}, {Milisavljevic}, {Miceli}, {Temim},
  {Rho}, {Nagataki}, {Ono}, {Sapienza}, \& {Greco}}]{2025A&A...696A.188O}
{Orlando}, S., {Janka}, H.-T., {Wongwathanarat}, A., {et~al.}
  2025{\natexlab{a}}, \aap, 696, A188

\bibitem[{{Orlando} {et~al.}(2025{\natexlab{b}}){Orlando}, {Janka},
  {Wongwathanarat}, {Dickinson}, {Milisavljevic}, {Miceli}, {Bocchino},
  {Temim}, {De Looze}, \& {Patnaude}}]{2025A&A...696A.108O}
{Orlando}, S., {Janka}, H.-T., {Wongwathanarat}, A., {et~al.}
  2025{\natexlab{b}}, \aap, 696, A108

\bibitem[{{Orlando} {et~al.}(2023){Orlando}, {Miceli}, {Lo Cicero}, \&
  {Ustamujic}}]{2023MmSAI..94a..13O}
{Orlando}, S., {Miceli}, M., {Lo Cicero}, U., \& {Ustamujic}, S. 2023, in
  Memorie della Societa Astronomica Italiana, Vol.~94, 13

\bibitem[{{Orlando} {et~al.}(2025{\natexlab{c}}){Orlando}, {Miceli}, {Ono},
  {Nagataki}, {Aloy}, {Bocchino}, {Gabler}, {Giudici}, {Giuffrida}, {Greco},
  {La Malfa}, {Lee}, {Obergaulinger}, {Petruk}, {Sapienza}, {Ustamujic}, \&
  {Weng}}]{2025A&A...699A.305O}
{Orlando}, S., {Miceli}, M., {Ono}, M., {et~al.} 2025{\natexlab{c}}, \aap, 699,
  A305

\bibitem[{{Orlando} {et~al.}(2019{\natexlab{a}}){Orlando}, {Miceli}, {Petruk},
  {Ono}, {Nagataki}, {Aloy}, {Mimica}, {Lee}, {Bocchino}, {Peres}, \&
  {Guarrasi}}]{2019A&A...622A..73O}
{Orlando}, S., {Miceli}, M., {Petruk}, O., {et~al.} 2019{\natexlab{a}}, \aap,
  622, A73

\bibitem[{{Orlando} {et~al.}(2015){Orlando}, {Miceli}, {Pumo}, \&
  {Bocchino}}]{2015ApJ...810..168O}
{Orlando}, S., {Miceli}, M., {Pumo}, M.~L., \& {Bocchino}, F. 2015, \apj, 810,
  168

\bibitem[{{Orlando} {et~al.}(2016){Orlando}, {Miceli}, {Pumo}, \&
  {Bocchino}}]{2016ApJ...822...22O}
{Orlando}, S., {Miceli}, M., {Pumo}, M.~L., \& {Bocchino}, F. 2016, \apj, 822,
  22

\bibitem[{{Orlando} {et~al.}(2020){Orlando}, {Ono}, {Nagataki}, {Miceli},
  {Umeda}, {Ferrand}, {Bocchino}, {Petruk}, {Peres}, {Takahashi}, \&
  {Yoshida}}]{2020A&A...636A..22O}
{Orlando}, S., {Ono}, M., {Nagataki}, S., {et~al.} 2020, \aap, 636, A22

\bibitem[{{Orlando} {et~al.}(2019{\natexlab{b}}){Orlando}, {Pillitteri},
  {Bocchino}, {Daricello}, \& {Leonardi}}]{2019RNAAS...3..176O}
{Orlando}, S., {Pillitteri}, I., {Bocchino}, F., {Daricello}, L., \&
  {Leonardi}, L. 2019{\natexlab{b}}, Research Notes of the American
  Astronomical Society, 3, 176

\bibitem[{{Orlando} {et~al.}(2022){Orlando}, {Wongwathanarat}, {Janka},
  {Miceli}, {Nagataki}, {Ono}, {Bocchino}, {Vink}, {Milisavljevic}, {Patnaude},
  \& {Peres}}]{2022A&A...666A...2O}
{Orlando}, S., {Wongwathanarat}, A., {Janka}, H.~T., {et~al.} 2022, \aap, 666,
  A2

\bibitem[{{Orlando} {et~al.}(2021){Orlando}, {Wongwathanarat}, {Janka},
  {Miceli}, {Ono}, {Nagataki}, {Bocchino}, \& {Peres}}]{2021A&A...645A..66O}
{Orlando}, S., {Wongwathanarat}, A., {Janka}, H.~T., {et~al.} 2021, \aap, 645,
  A66

\bibitem[{{Ravi} {et~al.}(2021){Ravi}, {Park}, {Zhekov}, {Miceli}, {Orlando},
  {Frank}, \& {Burrows}}]{2021ApJ...922..140R}
{Ravi}, A.~P., {Park}, S., {Zhekov}, S.~A., {et~al.} 2021, \apj, 922, 140

\bibitem[{{Ravi} {et~al.}(2024){Ravi}, {Park}, {Zhekov}, {Orlando}, {Miceli},
  {Frank}, {Broos}, \& {Burrows}}]{2024ApJ...966..147R}
{Ravi}, A.~P., {Park}, S., {Zhekov}, S.~A., {et~al.} 2024, \apj, 966, 147

\bibitem[{{Reed} {et~al.}(1995){Reed}, {Hester}, {Fabian}, \&
  {Winkler}}]{1995ApJ...440..706R}
{Reed}, J.~E., {Hester}, J.~J., {Fabian}, A.~C., \& {Winkler}, P.~F. 1995,
  \apj, 440, 706

\bibitem[{{Rho} {et~al.}(2024){Rho}, {Park}, {Arendt}, {Matsuura},
  {Milisavljevic}, {Temim}, {De Looze}, {Blair}, {Rest}, {Fox}, {Ravi}, {Koo},
  {Barlow}, {Burrows}, {Chevalier}, {Clayton}, {Fesen}, {Fransson}, {Fryer},
  {Gomez}, {Janka}, {Kirchschlager}, {Laming}, {Orlando}, {Patnaude}, {Pavlov},
  {Plucinsky}, {Posselt}, {Priestley}, {Raymond}, {Sartorio}, {Schmidt},
  {Slane}, {Smith}, {Sravan}, {Vink}, {Weil}, {Wheeler}, \&
  {Yoon}}]{2024ApJ...969L...9R}
{Rho}, J., {Park}, S.-H., {Arendt}, R., {et~al.} 2024, \apjl, 969, L9

\bibitem[{{Sapienza} {et~al.}(2024){Sapienza}, {Miceli}, {Bamba}, {Orlando},
  {Lee}, {Nagataki}, {Ono}, {Katsuda}, {Mori}, {Sawada}, {Terada}, {Giuffrida},
  \& {Bocchino}}]{2024ApJ...961L...9S}
{Sapienza}, V., {Miceli}, M., {Bamba}, A., {et~al.} 2024, \apjl, 961, L9

\bibitem[{{Sapienza} {et~al.}(2025){Sapienza}, {Miceli}, {Ono}, {Nagataki},
  {Yoshida}, {Greco}, {Orlando}, \& {Bocchino}}]{2025ApJ...990L...5S}
{Sapienza}, V., {Miceli}, M., {Ono}, M., {et~al.} 2025, \apjl, 990, L5

\bibitem[{{Sun} {et~al.}(2025){Sun}, {Orlando}, {Greco}, {Miceli}, {Chen},
  {Vink}, \& {Zhou}}]{2025ApJ...981...26S}
{Sun}, L., {Orlando}, S., {Greco}, E., {et~al.} 2025, \apj, 981, 26

\bibitem[{{Thorstensen} {et~al.}(2001){Thorstensen}, {Fesen}, \& {van den
  Bergh}}]{2001AJ....122..297T}
{Thorstensen}, J.~R., {Fesen}, R.~A., \& {van den Bergh}, S. 2001, \aj, 122,
  297

\bibitem[{{Vance} {et~al.}(2020){Vance}, {Young}, {Fryer}, \&
  {Ellinger}}]{2020ApJ...895...82V}
{Vance}, G.~S., {Young}, P.~A., {Fryer}, C.~L., \& {Ellinger}, C.~I. 2020,
  \apj, 895, 82

\bibitem[{{Weil} {et~al.}(2020){Weil}, {Fesen}, {Patnaude}, {Raymond},
  {Chevalier}, {Milisavljevic}, \& {Gerardy}}]{2020ApJ...891..116W}
{Weil}, K.~E., {Fesen}, R.~A., {Patnaude}, D.~J., {et~al.} 2020, \apj, 891, 116

\bibitem[{{Wongwathanarat} {et~al.}(2013){Wongwathanarat}, {Janka}, \&
  {M{\"u}ller}}]{2013A&A...552A.126W}
{Wongwathanarat}, A., {Janka}, H.~T., \& {M{\"u}ller}, E. 2013, \aap, 552, A126

\bibitem[{{Wongwathanarat} {et~al.}(2017){Wongwathanarat}, {Janka},
  {M{\"u}ller}, {Pllumbi}, \& {Wanajo}}]{2017ApJ...842...13W}
{Wongwathanarat}, A., {Janka}, H.-T., {M{\"u}ller}, E., {Pllumbi}, E., \&
  {Wanajo}, S. 2017, \apj, 842, 13

\bibitem[{{Wongwathanarat} {et~al.}(2015){Wongwathanarat}, {M{\"u}ller}, \&
  {Janka}}]{2015A&A...577A..48W}
{Wongwathanarat}, A., {M{\"u}ller}, E., \& {Janka}, H.-T. 2015, \aap, 577, A48

\bibitem[{{Xrism Collaboration} {et~al.}(2025){Xrism Collaboration}, {Audard},
  {Awaki}, {Ballhausen}, {Bamba}, {Behar}, {Boissay-Malaquin}, {Brenneman},
  {Brown}, {Corrales}, {Costantini}, {Cumbee}, {D{\'\i}az Trigo}, {Done},
  {Dotani}, {Ebisawa}, {Eckart}, {Eckert}, {Enoto}, {Eguchi}, {Ezoe}, {Foster},
  {Fujimoto}, {Fujita}, {Fukazawa}, {Fukushima}, {Furuzawa}, {Gallo},
  {Garc{\'\i}a}, {Gu}, {Guainazzi}, {Giuffrida}, {Hagino}, {Hamaguchi},
  {Hatsukade}, {Hayashi}, {Hayashi}, {Hell}, {Hodges-Kluck}, {Hornschemeier},
  {Ichinohe}, {Ishi}, {Ishida}, {Ishikawa}, {Ishisaki}, {Kaastra}, {Kallman},
  {Kara}, {Katsuda}, {Kanemaru}, {Kelley}, {Kikuchi}, {Kilbourne}, {Kitamoto},
  {Kobayashi}, {Kohmura}, {Kubota}, {Leutenegger}, {Loewenstein}, {Maeda},
  {Markevitch}, {Matsumoto}, {Matsushima}, {Matsushita}, {McCammon},
  {McNamara}, {Mernier}, {Miller}, {Miller}, {Mitsuishi}, {Mizumoto}, {Mizuno},
  {Mori}, {Mukai}, {Murakami}, {Mushotzky}, {Nakajima}, {Nakazawa}, {Ness},
  {Nobukawa}, {Nobukawa}, {Noda}, {Odaka}, {Ogawa}, {Ogorzalek}, {Okajima},
  {Ota}, {Paltani}, {Petre}, {Plucinsky}, {Porter}, {Pottschmidt}, {Sato},
  {Sato}, {Sawada}, {Seta}, {Shidatsu}, {Shimoda}, {Simionescu}, {Smith},
  {Suzuki}, {Szymkowiak}, {Takahashi}, {Takeo}, {Tamagawa}, {Tamura}, {Tanaka},
  {Tanimoto}, {Tashiro}, {Terada}, {Terashima}, {Tsuboi}, {Tsujimoto},
  {Tsunemi}, {Tsuru}, {Uchida}, {Uchida}, {Uchida}, {Uchiyama}, {Ueda}, {Ueda},
  {Uno}, {Vink}, {Watanabe}, {Williams}, {Yamada}, {Yamada}, {Yamaguchi},
  {Yamaoka}, {Yamasaki}, {Yamauchi}, {Yamauchi}, {Yaqoob}, {Yoneyama},
  {Yoshida}, {Yukita}, {Zhuravleva}, {Miceli}, \&
  {Sapienza}}]{2025PASJ...77S.193X}
{Xrism Collaboration}, {Audard}, M., {Awaki}, H., {et~al.} 2025, \pasj, 77,
  S193

\end{thebibliography}

%

\end{document}